# SS-ZKR: Spatial-Semantic Zero-Knowledge Routing for Privacy-Preserving Multi-Agent Collaboration

*Hassan Touheed*
*htouheed@hotmail.com*

**Abstract**

Foundational agent interoperability standards, notably the Agent-to-Agent (A2A) protocol for peer collaboration and the Model Context Protocol (MCP) for tool integration, have advanced multi-agent system (MAS) communication significantly, and complementary identity frameworks leveraging W3C Decentralised Identifiers (DIDs) and Verifiable Credentials (VCs) provide cryptographic agent authentication. However, no existing protocol supports content-based semantic routing of agent payloads across organisational trust boundaries without requiring the routing intermediary to decrypt the payload, which is a hard constraint in compliance-sensitive environments governed by GDPR, HIPAA, and MiFID II. We propose SS-ZKR, a three-mechanism privacy-preserving routing protocol designed as a complementary layer atop A2A/MCP. Mechanism I introduces blind routing via differentially private semantic intent vectors cryptographically bound to zero-knowledge proofs of payload-schema consistency. Mechanism II offers vector-weighted adaptive payload sanitisation with formal (ε, δ)-differential privacy guarantees for numerical fields and heuristic semantic aggregation for textual fields. Mechanism III presents a spatial-to-cryptographic policy compiler that translates visually defined trust-zone topologies into deterministic ZK access circuits.

We provide a formal threat model, analyse information leakage bounds of intent vectors, present pseudocode for all three mechanisms, and give analytical complexity comparisons against TEE-based and homomorphic encryption-based routing baselines. SS-ZKR lets enterprises in financial services, healthcare, and defence orchestrate heterogeneous AI agents across regulatory boundaries without exposing proprietary data to routing infrastructure.

## I. INTRODUCTION

The transition from isolated Large Language Models (LLMs) to autonomous, goal-oriented AI agents has catalysed the emergence of the Internet of Agents (IoA) paradigm [1], [2]. Gartner reported a 1,445% increase in multi-agent system (MAS) inquiries between Q1 2024 and Q2 2025 [3], and industry projections place the agentic AI market at $52 billion by 2030 [4]. Standardisation has advanced in parallel: Google's Agent-to-Agent (A2A) protocol [5], now governed by the Linux Foundation with 150+ supporting organisations, provides agent discovery via Agent Cards, task lifecycle management, and collaboration primitives. Anthropic's Model Context Protocol (MCP) [6] standardises agent-to-tool integration. Decentralised identity frameworks based on W3C DIDs [7] and VCs [8] provide cryptographic agent authentication [9]–[11].

Despite this progress, a critical routing-layer gap persists. In enterprise environments governed by strict data sovereignty mandates (GDPR, HIPAA, MiFID II, ITAR), the following constraint applies: **no intermediary infrastructure (including internal routing middleware) may access the plaintext** **content of inter-agent payloads without explicit authorisation for each data class contained therein.** Current agent communication protocols implicitly assume plaintext payload access at the routing layer. A2A's task messages are transmitted in cleartext within the secure channel; MCP tool invocations expose parameters to the host. Neither protocol provides a mechanism for content-based routing decisions on encrypted payloads.

This constraint creates a **compliance deadlock** that, as articulated by enterprise architects and compliance officers in industry analyses [3], [15] and observed in practitioner reports across regulated sectors, is already impeding MAS adoption in regulated industries:

- In healthcare, a clinical AI agent cannot query a billing AI agent through shared middleware without exposing Protected Health Information (PHI) to the middleware layer (HIPAA §164.502).
- In financial services, a risk assessment agent cannot query a trading surveillance agent via a central orchestrator without exposing position data to the routing layer (MiFID II Article 16, Chinese wall requirements).
- In defence, classified AI agents cannot expose context windows to unclassified routing infrastructure (NIST SP 800-53, multi-level security requirements).
- In cross-border operations, a European agent collaborating with a US-based agent through a routing layer in a third jurisdiction triggers GDPR Chapter V transfer mechanism requirements if the payload is decrypted at the routing point.

This paper proposes the **Spatial-Semantic Zero-Knowledge Routing (SS-ZKR)** protocol to address this gap. SS-ZKR is designed not as a replacement for A2A or MCP, but as a *complementary privacy-preserving routing layer* that extends their capabilities into compliance-sensitive environments. Our contributions are:

1) **C1:** A formal threat model for multi-agent routing intermediaries, distinguishing semi-honest and malicious adversary models (Section III).
2) **C2:** A blind routing mechanism using differentially private semantic intent vectors cryptographically bound to zk-SNARKs of payload-schema consistency, with formal analysis of information leakage bounds (Section IV-A).
3) **C3:** A vector-weighted payload sanitisation model with $(\varepsilon, \delta)$-differential privacy guarantees for graduated response obfuscation (Section IV-B).
4) **C4:** A spatial-to-cryptographic policy compiler with formal grammar specification for visual trust-zone definition (Section IV-C).
5) **C5:** Analytical complexity comparison against TEE-based and homomorphic encryption-based routing baselines (Section VI).

# II. RELATED WORK

### *A. Agent Communication Protocols*

The A2A protocol [5] defines a client-server interaction model between opaque agentic applications. Agents advertise capabilities via JSON Agent Cards, interactions follow a task lifecycle (created → in-progress → completed), and authentication uses OpenAPI-like schemes. Version 0.3 [12] added gRPC transport and signed security cards. MCP [6] standardises agent-to-tool integration through a complementary client-server architecture. Adimulam et al. [13] provide a comprehensive analysis of how A2A and MCP form the dual communication substrate for orchestrated MAS. Neither protocol addresses privacy-preserving routing: A2A task messages and MCP tool invocations transmit payload content in plaintext within the secure channel.

### *B. Decentralised Identity for AI Agents*

Huang et al. [9] proposed a zero-trust identity framework for agentic AI using DIDs, VCs, ZKPs, and an Agent Naming Service (ANS), analysed against the MAESTRO threat model. The LOKA Protocol [10] introduced a Universal Agent Identity Layer (UAIL) with intent-centric communication and a Decentralised Ethical Consensus Protocol. Rodriguez Garzon et al. [11] demonstrated cross-domain agent authentication using DIDs/VCs over A2A with LangChain and AutoGen agents, revealing that delegating security orchestration to LLMs is unreliable. Fanitabasi [14] combined CQRS, blockchain, and ZKPs for MAS with a Decentralised Discovery Facility. These frameworks address identity verification (‘who is this agent?’) but do not address the routing question (‘given an encrypted payload, which agent should receive it and how much should they see?’).

### *C. Agent Routing Architectures*

Fauscette [15] articulated the Agentic Service Bus (ASB) concept, distinguishing ‘Soft A2A’ (negotiation via LLM reasoning) from ‘Hard A2A’ (execution via schema-enforced payloads). Microsoft’s Azure AI Foundry Agent Service [16] provides enterprise-grade agent infrastructure with Entra identity and A2A/MCP support. IBM’s BeeAI framework [17] and the AGNTCY project [18] offer open-source orchestration with federated agent directories. All require plaintext payload access for routing decisions.

### *D. Privacy-Preserving Search and Routing*

Encrypted search and routing have been studied extensively outside the MAS domain. Searchable Symmetric Encryption (SSE) [19] enables keyword search over encrypted data but is limited to exact-match queries, not semantic similarity. Functional Encryption (FE) [20] allows computation of specific functions on encrypted inputs but introduces prohibitive computational overhead for high-dimensional vector operations. Private Information Retrieval (PIR) [21] enables database queries without revealing the query, but scales poorly with database size. Secure Multi-Party Computation (SMPC) [22] can compute similarity functions jointly but requires interactive protocols incompatible with stateless routing. **None of these approaches have been applied to semantic capability matching in federated MAS environments**, which requires approximate nearest-neighbour search over high-dimensional encrypted vectors at low latency. SS-ZKR addresses this gap.

### *E. Embedding Information Leakage*

A critical consideration for any embedding-based routing scheme is information leakage. Morris et al. [23] demonstrated that text embeddings reveal almost as much information as the original text, achieving high-accuracy text reconstruction from embeddings. This finding directly impacts any system that transmits semantic embeddings to a potentially untrusted intermediary. SS-ZKR addresses this through differentially private embedding perturbation (Section IV-A), a mitigation not present in any existing MAS routing proposal.

**TABLE I:** Comparative analysis of existing protocols against privacy-preserving routing requirements

| Framework | Agent Discovery | Task Mgmt | Authn | Encrypted Routing | Adaptive Sanitise | Visual Policy | DP Guarantee |
|---|---|---|---|---|---|---|---|
| A2A [5] | ✓ | ✓ | ✓ | ✗ | ✗ | ✗ | ✗ |
| MCP [6] | Partial | ✗ | Partial | ✗ | ✗ | ✗ | ✗ |
| Huang [9] | ✓ | ✗ | ✓ | ✗ | ✗ | ✗ | ✗ |
| LOKA [10] | ✓ | Partial | ✓ | ✗ | ✗ | ✗ | ✗ |
| ASB [15] | ✓ | ✓ | Partial | ✗ | ✗ | ✗ | ✗ |
| **SS-ZKR** | ✓* | ✓* | ✓* | ✓ | ✓ | ✓ | ✓ |

**Note:** * SS-ZKR inherits Agent Discovery, Task Management, and Authentication from underlying A2A/MCP/DID infrastructure. SS-ZKR is a complementary routing layer that adds encrypted routing, adaptive sanitisation, visual policy compilation, and DP guarantees—it is not a standalone protocol.

## III. THREAT MODEL AND SECURITY DEFINITIONS

### *A. System Model*

The system comprises three classes of entity: (1) **Source agents** (S) that initiate capability requests; (2) **Destination agents** (D) that provide capabilities; and (3) the **Agentic Mesh Fabric** (AMF), a stateless routing intermediary that performs capability matching and payload forwarding. Agents authenticate via DIDs and present VCs to establish trust levels. Communication between agents and the AMF uses TLS 1.3. The AMF maintains a **Semantic Capability Registry** (SCR) populated from A2A Agent Cards, containing capability vectors $\{v_c\}$ but no agent payload data.

### *B. Adversary Model*

We consider two adversary classes, informed by the IoA threat taxonomy of Wang et al. [26]:

**Adversary A1 (Semi-honest AMF):** The routing intermediary follows the protocol correctly but attempts to infer information about agent payloads from observable routing metadata. This models a curious infrastructure provider, a compromised but functional middleware layer, or an insider threat with read access to routing logs. A1 observes: intent vectors $v_i$, ZK proofs $\pi_i$, encrypted payload ciphertexts E(payload), routing decisions, and timing metadata.

**Adversary A2 (Malicious agent):** A participating agent that deviates from the protocol to: (a) spoof capabilities via false Agent Cards (registration spoofing); (b) declare benign intent vectors while transmitting adversarial payloads (intent-payload mismatch); or (c) attempt to extract sensitive data from destination agents via carefully crafted queries (data exfiltration via semantic probing).

### *C. Security Goals*

SS-ZKR aims to provide the following guarantees:

**G1 (Routing Privacy):** Adversary A1 cannot determine the semantic content of a routed payload with probability significantly greater than random guessing, formalised as $(\varepsilon, \delta)$-differential privacy on the intent vector.

**G2 (Intent-Payload Integrity):** Adversary A2 cannot successfully route a payload whose semantic content deviates from the declared intent vector, enforced via zk-SNARK verification.

**G3 (Graduated Disclosure):** Destination agents return response data at a granularity proportional to the verified trust level of the requester, with $(\varepsilon, \delta)$-DP guarantees on the sanitisation function.

**G4 (Policy Consistency):** The compiled ZK access policies are semantically equivalent to the administrator-defined spatial trust-zone topology.

### *D. Explicit Non-Goals*

SS-ZKR does not protect against: (a) compromise of the source or destination agent's local runtime (endpoint security is orthogonal); (b) traffic analysis attacks based on message timing, size, or frequency (requires orthogonal padding/mixing defences); (c) collusion between a malicious agent and a compromised AMF (this collapses to full system compromise); (d) denial-of-service attacks on the routing layer.

## IV. THE SS-ZKR PROTOCOL

### *A. Mechanism I: Differentially Private Blind Routing*

The core challenge of blind routing is enabling the AMF to perform semantic capability matching without accessing the payload. Naïvely transmitting the intent vector $v_i$ to the AMF leaks information: Morris et al. [23] demonstrated that text embeddings can be inverted to reconstruct approximate plaintext. We address this through a four-step protocol:

**Step 1 — Intent Embedding.** The source agent S generates a d-dimensional semantic intent vector $v_i \in \mathbb{R}^d$ from its capability request using a shared embedding model $f_{embed}$ agreed upon at mesh registration (analogous to the schema agreement in A2A Agent Cards). The model is distributed as a hash-verified container image; updates require multi-party approval from participating organisations and trigger re-embedding of all capability vectors in the SCR.

**Step 2 — Differential Privacy Perturbation.** To mitigate embedding inversion attacks, S samples Gaussian noise $\eta \sim N(0, \sigma^2 I_d)$ and applies it to the intent vector. Because cosine similarity assumes unit-normalised vectors and raw additive Gaussian noise moves the perturbed vector off the unit hypersphere (with expected magnitude $\sqrt{(1 + d\sigma^2)}$), we re-normalise after perturbation:

$$\tilde{v}_i = (v_i + \eta) / \|v_i + \eta\|, \quad \text{where } \eta \sim N(0, \sigma^2 I_d)$$

where $\sigma$ is calibrated to achieve $(\varepsilon, \delta)$-differential privacy via the Gaussian mechanism [24]. By the post-processing property of differential privacy, the re-normalisation step preserves the $(\varepsilon, \delta)$-DP guarantee. We note that additive Gaussian noise followed by L2 re-normalisation is a pragmatic approach; a theoretically cleaner alternative is the von Mises-Fisher (vMF) mechanism, which samples noise directly on the unit hypersphere and preserves directional statistics intrinsically. Adopting the vMF mechanism requires re-deriving the sensitivity bounds under the concentration parameter $\kappa$ rather than $\sigma$, and is identified as a refinement for the prototype implementation (Section VII). The privacy-utility tradeoff is governed by ε: lower ε provides stronger privacy but reduces routing accuracy (cosine similarity with the true capability vector). We define the **routing accuracy** as the probability that the perturbed vector $\tilde{v}_i$ matches the correct capability vector $v_{c^*}$ when $\mathrm{sim}(\tilde{v}_i, v_{c^*}) \geq \theta$. Section VI gives an analytical characterisation of this tradeoff.

**Step 3 — Zero-Knowledge Schema Attestation with Noise Binding.** S constructs a zk-SNARK proof $\pi_i$ attesting a relation that cryptographically binds the original intent vector, the noise sample, the perturbed vector sent to the AMF, and the encrypted payload. The relation is:

$$R(\pi_i) : \exists\ (v_i, \eta, \text{payload}) \text{ s.t.}$$

$$\tilde{v}_i = (v_i + \eta) / \|v_i + \eta\| \quad \text{[noise binding]}$$

$$\wedge\ \mathrm{Schema}(\text{payload}, C(v_i)) = 1 \quad \text{[schema compliance]}$$

$$\wedge\ E_{pk}(\text{payload}) = \text{ciphertext} \quad \text{[payload binding]}$$

The circuit takes the original intent vector $v_i$, the noise sample η, and the plaintext payload as *private* witnesses, and the perturbed vector $\tilde{v}_i$, schema class $C(v_i)$, and ciphertext as *public* inputs. $C(v_i)$ is the schema class associated with the intent vector, Schema(·, ·) is the schema validation predicate, and $E_{pk}(\cdot)$ is the encryption function. The noise-binding clause is essential: without it, Adversary A2 could prove schema compliance against one intent vector $v_i$ while transmitting a perturbed vector $\tilde{v}_j$ derived from a different intent $v_j$, tricking the AMF into routing the payload to an unauthorised destination. With the binding, the AMF's verification of $\pi_i$ confirms that the exact $\tilde{v}_i$ used for routing is cryptographically tied to the intent vector whose schema class was attested (Goal G2). We assume the Groth16 proving system for concrete latency estimates, noting that alternative schemes (PLONK, Halo2) trade proving time for reduced trust assumptions by eliminating the per-circuit trusted setup. The re-normalisation constraint (division and L2 norm computation) adds approximately $10^4$ additional R1CS (Rank-1 Constraint System) constraints to the circuit—the standard complexity metric for Groth16 and PLONK proving systems; recent advances in SNARK-friendly arithmetic make this tractable.

The schema class identifier $C(v_i)$ is transmitted alongside the proof as a coarse-grained routing hint. This represents a controlled information disclosure: schema classes are broad capability categories (e.g., 'incident-correlation', 'financial-reporting') designed to be non-sensitive. The DP perturbation on the intent vector provides fine-grained privacy *within* the schema class, ensuring that while the AMF learns the general category of a request, it cannot determine its specific content or the sensitive data it references.

**Step 4 — Blind Route Execution.** The AMF receives the tuple ($\tilde{v}_i$, $\pi_i$, ciphertext). It: (a) verifies $\pi_i$; (b) computes $\mathrm{sim}(\tilde{v}_i, v_c)$ against all capability vectors in the SCR; (c) routes the ciphertext to the destination agent D whose capability vector maximises similarity above threshold θ. The AMF never decrypts the payload.

**Algorithm 1:** Differentially Private Blind Routing

Input: request R, embedding model f_embed, privacy parameter ε, threshold θ
Output: routing decision (destination agent D* or ⊥)
 1: S: v_i ← f_embed(R.intent)

```
 2: S: σ ← Δf · √(2 ln(1.25/δ)) / ε          // Gaussian mechanism
 3: S: sample η ~ N(0, σ²I_d)
 4: S: ṽ_i ← (v_i + η) / ||v_i + η||          // perturb and re-normalise
 5: S: ciphertext ← E_pk(R.payload)
 6: S: π_i ← zkSNARK.Prove(
        witness: (v_i, η, R.payload);
        public:  (ṽ_i, C(v_i), ciphertext))          // noise-bound proof
 7: S → AMF: (ṽ_i, C(v_i), π_i, ciphertext)
 8: AMF: if zkSNARK.Verify(π_i) = false: return ⊥
 9: AMF: D* ← argmax_{D} sim(ṽ_i, v_c^D) s.t. sim > θ
10: AMF: if D* = ∅: return ⊥
11: AMF → D*: (ciphertext, ṽ_i, S.DID, S.VCs)
12: return D*
```

---

### *B. Mechanism II: Vector-Weighted Adaptive Sanitisation*

Upon receiving a routed request, the destination agent D must determine the granularity of its response based on the requester's trust level. Static DLP (regex-based PII detection) is insufficient for semantically rich agent payloads where sensitivity is context-dependent. SS-ZKR introduces vector-weighted sanitisation with formal privacy guarantees.

**Definition 1 (Semantic Trust Distance).** *Let $v_{auth}$ denote the requester's authorised access vector (derived from its verified VCs and the compiled trust-zone policy), and $v_{sec}$ denote the destination's security classification vector (derived from its data sensitivity labels registered in the SCR). The semantic trust distance is:*

$$D(S, D) = 1 - \mathrm{sim}(v_{auth}, v_{sec})$$

**Definition 2 (Sanitisation Function).** *The sanitisation function Φ maps a raw response r and a trust distance D to a sanitised response r′:*

$$r' = \Phi(r, D(S, D)) = M_\varepsilon(\mathrm{Agg}(r, D(S, D)))$$

where $\mathrm{Agg}(\cdot, \cdot)$ is a semantic aggregation function (parameterised by D) that reduces specificity (figures → ranges, entities → roles, timestamps → periods), and $M_\varepsilon$ is the (ε, δ)-DP mechanism applied to the aggregated output. The access vector v*auth* is computed by applying f*embed* to the concatenation of the requester's VC attribute labels, producing a representation in the same vector space as capability vectors. This provides **Goal G3**: the composition of semantic aggregation and DP noise ensures that even if an adversary obtains multiple sanitised responses at different trust levels, the intersection of those responses does not reveal the raw data beyond the DP bound. For numerical response fields (e.g., financial figures, performance metrics), the Gaussian mechanism provides formal (ε, δ)-DP guarantees. For textual fields, formal DP is substantially harder: mechanisms such as the Exponential Mechanism applied to vocabulary logits can provide formal DP for token-level generation but severely degrade utility in practice. We therefore adopt a hybrid approach: (i) formal (ε, δ)-DP on numerical fields via the Gaussian mechanism; (ii) heuristic semantic sanitisation on textual fields via the Agg function, which reduces specificity proportional to trust distance. We explicitly label the textual sanitisation as heuristic rather than claiming formal DP guarantees on generated text—a principled treatment of formal DP for free-form synthesis is an open research problem and is acknowledged in Section VII.

**Security of the Sanitisation Model.** The synthesis model used for semantic aggregation operates locally within the destination agent's trust boundary. It is not exposed to external agents and is subject to the destination's own security controls. If the synthesis model is compromised (e.g., via model poisoning), the exposure is limited to the destination agent's own data—a risk that exists independently of SS-ZKR and is addressed by endpoint security measures outside our scope (see Section III-D).

**Algorithm 2:** Vector-Weighted Adaptive Sanitisation

```
Input: source agent S, raw response r, trust-zone policy P
Output: sanitised response r′
 1: v_auth ← f_embed(concat(S.VC_attributes))
 2: v_sec  ← D.securityClassification
 3: D(S,D) ← 1 − cosineSim(v_auth, v_sec)
 4: τ      ← P.sanitisationThreshold(zone(S), zone(D))
 5: if D(S,D) ≤ τ:
 6:     r_agg ← r                        // full granularity
 7: else:
 8:     r_agg ← Agg(r, D(S,D))           // semantic aggregation
 9: σ_san  ← Δf_san · √(2 ln(1.25/δ)) / ε_san
10: numFields(r_agg) ← r_agg.numericalFields + N(0, σ_san²)   // formal (ε,δ)-DP via Gaussian mechanism
11: txtFields(r_agg) ← Generalise(r_agg.textualFields, level(D(S,D)))   // heuristic semantic aggregation
12: r′ ← assemble(numFields, txtFields)
13: return r′
```

### *C. Mechanism III: Spatial-to-Cryptographic Policy Compiler*

The compiler translates visually defined trust-zone topologies into deterministic ZK access circuits. We define the visual policy language formally:

**Definition 3 (Trust-Zone Topology).** *A trust-zone topology T is a tuple $(N, Z, \subseteq, \kappa)$ where: N is a set of agent nodes; Z is a set of trust zones (closed regions in $\mathbb{R}^2$); $\subseteq : N \to P(Z)$ maps each agent to the set of zones containing it; and $\kappa : Z \times Z \to [0, 1]$ defines the trust coefficient between zone pairs.*

**Definition 4 (Compiled Policy).** *For each ordered pair of zones $(Z_i, Z_j)$ where $Z_i \neq Z_j$, the compiler emits: (a) a ZK circuit $C_{i,j}$ encoding the VC attestations an agent in $Z_i$ must prove to route a payload to an agent in $Z_j$; (b) a sanitisation threshold $\tau_{i,j} = 1 - \kappa(Z_i, Z_j)$; and (c) an AMF routing rule $R_{i,j}$ specifying the privacy parameter $\varepsilon_{i,j}$ for the DP perturbation of intent vectors crossing this boundary.*

The compilation pipeline proceeds in three phases:

1) **Topological parsing:** The canvas is discretised into a planar subdivision. Zone containment and intersection relationships are extracted using computational geometry (point-in-polygon tests, polygon intersection detection). The output is a zone adjacency graph $G_Z$.

2) **Trust coefficient derivation:** For adjacent zones, κ is derived from zone nesting depth (nested zones share higher trust) and administrator-specified overrides. For non-adjacent zones, κ is computed as the minimum κ along the shortest path in $G_Z$.

3) **Circuit synthesis:** Each boundary crossing is compiled to an arithmetic circuit encoding the conjunction of required VC attribute proofs. The circuit complexity is bounded by $O(a \cdot k)$ where a is the number of required attributes and k is the maximum VC chain depth.

**Goal G4 (Policy Consistency)** requires that the compiled policies faithfully represent the spatial topology. We outline a verification approach via bi-simulation: for every pair of agents (S, D) and every payload, the routing decision under the compiled policy should be identical to the decision that a hypothetical omniscient oracle would make by evaluating the spatial topology directly. Formal proof of this property, and the specification of the bi-simulation relation in detail, is identified as critical future work (Section VII).

**Algorithm 3:** Spatial-to-Cryptographic Policy Compilation

```
Input: canvas topology T = (N, Z, ⊆, κ_overrides)
Output: compiled policy set {(C_{i,j}, τ_{i,j}, ε_{i,j})} for all zone pairs
 1: G_Z ← buildAdjacencyGraph(Z)      // point-in-polygon, intersection
 2: for each zone pair (Z_i, Z_j) in G_Z:
 3:   if Z_j ⊂ Z_i:                    // nested zones
 4:     κ(Z_i, Z_j) ← 0.9              // high default trust
 5:   elif Z_i ∩ Z_j ≠ ∅:              // overlapping zones
 6:     κ(Z_i, Z_j) ← 0.5              // moderate default trust
 7:   else:                            // non-adjacent
 8:     κ(Z_i, Z_j) ← minPathκ(G_Z, Z_i, Z_j)  // transitive decay
 9:   if κ_overrides(Z_i, Z_j) exists:
10:     κ(Z_i, Z_j) ← κ_overrides(Z_i, Z_j)
11:   τ_{i,j} ← 1 − κ(Z_i, Z_j)        // sanitisation threshold
12:   ε_{i,j} ← ε_base / κ(Z_i, Z_j)   // stricter DP at lower trust
13:   attrs ← requiredVCAttributes(Z_i, Z_j)
14:   C_{i,j} ← synthesizeZKCircuit(attrs) // O(a · k) constraints
15: return {(C_{i,j}, τ_{i,j}, ε_{i,j})}
```

# V. ENTERPRISE DEPLOYMENT: THE AGENTIC MESH FABRIC

## *A. Architecture*

The Agentic Mesh Fabric (AMF) is the deployment architecture for SS-ZKR. It operates as a stateless privacy layer between agents, complementing the Agentic Service Bus pattern [15] with cryptographic routing guarantees. The AMF:

- **Delegates communication semantics** to A2A (task lifecycle, Agent Cards, collaboration primitives).
- **Delegates tool integration** to MCP (function invocations, data source access).
- **Delegates identity management** to DID/VC infrastructure [9], [10].
- **Adds privacy-preserving routing** via SS-ZKR (blind routing, adaptive sanitisation, compiled access policies).

## *B. Topology Model*

The AMF supports three deployment topologies, corresponding to concentric trust zones in the spatial compiler:

**Private Mesh:** Internal deployment connecting an organisation's heterogeneous vendor-specific agents (e.g., Atlassian Rovo, Microsoft Copilot, ServiceNow AI). The AMF enforces internal clearance levels via vector-weighted sanitisation. Intent vector DP perturbation may be relaxed ($\varepsilon \to \infty$) within a single trust zone if the organisation accepts the information leakage risk internally.

**Hybrid Mesh:** When an internal agent requires external capabilities, requests are routed through a Governance Sidecar—a boundary proxy executing the full SS-ZKR protocol. The sidecar's behaviour is determined by the compiled ZK policies for the relevant zone boundary.

**Public Mesh:** Fully federated deployment across organisations. All three SS-ZKR mechanisms are active. This topology supports the emerging B2B agentic economy (exemplified by Google's AI Agent Marketplace and Microsoft's Foundry Agent Service), where organisations expose specialised AI capabilities as commercial services. Privacy-preserving routing is a prerequisite for competitive collaboration: organisations will not expose proprietary data to marketplace intermediaries.

### *C. Scenario: Cross-Platform Incident Resolution in Financial Services*

A Tier-1 bank operates under MiFID II Chinese wall requirements. Its Risk Assessment agent (Zone: Trading Floor, Classification: Restricted) detects anomalous patterns requiring correlation with its Compliance Monitoring agent (Zone: Compliance, Classification: Confidential) and an external Regulatory Reporting agent operated by a RegTech provider (Zone: External, Classification: Public).

Without SS-ZKR, this workflow is blocked: the internal routing layer cannot see trading data (Chinese wall), and the external RegTech provider's agent cannot receive un-sanitised compliance data (GDPR, MiFID II). With SS-ZKR:

1) The Risk Assessment agent generates a DP-perturbed intent vector ($\varepsilon = 2.0$ for internal routing, where leakage risk is partially accepted) and a zk-SNARK attesting payload schema compliance. The AMF routes to the Compliance agent without seeing trading position data.

2) The Compliance agent's response is sanitised via the vector-weighted model: full granularity to the Risk Assessment agent (same trust zone), aggregated to the external RegTech agent (cross-zone, $\tau = 0.7$).

3) The cross-boundary request to the RegTech agent uses strict DP perturbation ($\varepsilon = 0.5$) on the intent vector, ensuring the external AMF relay cannot reconstruct the query semantics.

The spatial compiler ensures that if a new regulatory zone is defined (e.g., a DORA compliance zone for digital operational resilience), the ZK access policies automatically recompile to enforce the new boundary without manual cryptographic configuration.

## VI. ANALYTICAL EVALUATION

### *A. Information Leakage Bounds for Intent Vectors*

We characterise the privacy-utility tradeoff of the DP perturbation. Let the embedding dimensionality $d = 768$ (typical for sentence transformers) and the intent vector sensitivity $\Delta f = \max \|v_i - v_i'\| = 2$ (normalised unit vectors). The Gaussian mechanism achieves $(\varepsilon, \delta)$-DP with $\sigma = \Delta f\sqrt{2 \ln(1.25/\delta)}/\varepsilon$. For $\varepsilon = 1.0$ and $\delta = 10^{-5}$:

$$\sigma \approx 2 \times \sqrt{2 \ln(125000)} / 1.0 \approx 2 \times 4.82 \approx 9.64$$

At this noise level, the cosine similarity between the perturbed and true intent vector is approximately $E[\mathrm{sim}(\tilde{v}_i, v_i)] = 1 / (1 + d\sigma^2/\|v_i\|^2)$, which for $d = 768$ and $\|v_i\| = 1$ yields $\approx 0.000014$. At $\varepsilon = 5.0$, $\sigma \approx 1.93$ and $E[\mathrm{sim}] \approx 0.00035$; at $\varepsilon = 10.0$, $\sigma \approx 0.96$ and $E[\mathrm{sim}] \approx 0.0014$. These absolute similarity values are deliberately low: a single global $\varepsilon$ chosen to satisfy strict cross-boundary privacy is, by construction, incompatible with high-utility routing in high dimensions. **This is precisely the motivation for the spatial compiler's per-boundary ε emission** (Section IV-C, Algorithm 3, line 12): rather than committing to a single $\varepsilon$ that must satisfy the worst-case adversary, SS-ZKR emits a distinct $\varepsilon_{\{i,j\}}$ per zone-pair boundary, calibrated to the trust coefficient $\kappa(Z_i, Z_j)$. Internal-trust-domain interactions where leakage to the routing layer is acceptable use relaxed $\varepsilon$ (e.g., 5–10), preserving routing utility; cross-organisational interactions use strict $\varepsilon$ (e.g., 0.5–2.0), accepting reduced utility in exchange for formal privacy guarantees against external observation. The sensitivity bound $\Delta f = 2$ used above corresponds to the worst-case L2 distance between any two unit-normalised intent vectors and is conservative; tighter sensitivity bounds derived from declared schema classes can yield lower noise in practice. The correct per-boundary evaluation metric is **top-k retrieval accuracy** against the destination registry, not absolute similarity. Although noise depresses absolute cosine similarity, the relative ranking of destination agents is preserved with non-trivial probability when noise is isotropic and capability vectors are well-separated in the embedding space: standard concentration arguments suggest that for moderate noise the correct destination remains within the top-k for k logarithmic in registry size n, with secondary disambiguation via schema matching and VC compatibility resolving the final routing decision. We acknowledge this is an analytical argument requiring empirical confirmation. Lower-dimensional intent vectors purpose-built for routing ($d = 32$ or $d = 64$) further improve the tradeoff: at $d = 64$ and $\varepsilon = 5.0$, $E[\mathrm{sim}] \approx 0.004$, a 12× improvement over $d = 768$. Empirical characterisation of top-k retrieval accuracy across varying $\varepsilon$, d, and registry sizes n is a primary objective of the planned prototype evaluation (Section VII). **The key insight is that the privacy-utility tradeoff is configurable per trust-zone boundary** via the spatial compiler: strict $\varepsilon$ at external boundaries, relaxed $\varepsilon$ internally.

### *B. Complexity Comparison Against Baselines*

**TABLE II:** Analytical complexity comparison of routing approaches

| Approach | Routing Latency | Privacy Guarantee | Trust Model | Deployment |
|---|---|---|---|---|
| Plaintext A2A | O(n) similarity | None | Trusted router | Simple |
| TEE (SGX/TDX) | O(n) + attestation | HW-based isolation | Trusted HW vendor | HW-dependent |
| HE-based (CKKS) | O(n · d) HE-mul | Computational | Semi-honest | Very high cost |
| SS-ZKR | O(n) + ZKP verify | (ε,δ)-DP + ZKP | Semi-honest AMF | Software-only (commodity cloud) |

SS-ZKR occupies a software-only middle ground: it provides formally characterised $(\varepsilon, \delta)$-DP privacy guarantees without the specialised hardware dependency of TEE approaches (Intel SGX, AMD SEV, ARM TrustZone) or the prohibitive computational cost of fully homomorphic operations, enabling deployment on commodity cloud infrastructure. The primary cost is the zk-SNARK proving time (estimated at 2–30 seconds for circuits of $10^5$–$10^6$ R1CS constraints using Groth16 on commodity hardware). This is acceptable for asynchronous agent workflows but prohibitive for real-time synchronous negotiation. Ongoing advances in GPU-accelerated proving and recursive SNARK composition are expected to reduce this to sub-second latency within 2–3 years [25].

## VII. LIMITATIONS AND FUTURE WORK

We identify the following limitations and corresponding research directions:

1) **Prototype implementation.** This paper presents a conceptual architecture with analytical evaluation. A proof-of-concept implementation using Circom/SnarkJS integrated with the A2A Python SDK is the immediate next step.

2) **Formal compiler verification.** The spatial-to-cryptographic compiler's correctness (Goal G4) is stated but not formally proved. Bi-simulation verification or model checking is required.

3) **Embedding model governance.** The shared embedding model is a centralisation vector. Techniques for cross-model intent vector alignment (e.g., via Procrustes alignment or contrastive learning) must be investigated.

4) **Composition attacks on sanitisation.** While the DP mechanism provides formal bounds on individual queries, the composition theorem applies to repeated queries. Tracking and enforcing a global privacy budget across multiple interactions is a systems challenge.

5) **User study for spatial compiler.** The usability claim requires empirical validation via a controlled study with compliance officers and enterprise architects.

6) **Formal DP on textual synthesis.** The sanitisation function provides formal $(\varepsilon, \delta)$-DP on numerical response fields but only heuristic privacy on textual fields produced by semantic aggregation. Extending formal DP guarantees to free-form text generation (for example, via the Exponential Mechanism over vocabulary logits with utility-preserving calibration, or DP-SGD-trained synthesis models with formal per-token privacy budgets) is an open research direction that would close this gap.

7) **Directional noise mechanisms.** Our current construction uses additive Gaussian noise followed by L2 re-normalisation. A theoretically cleaner alternative is the von Mises-Fisher mechanism, which samples noise directly on the unit hypersphere and preserves directional statistics intrinsically. Re-deriving the sensitivity analysis and privacy-utility tradeoff bounds under vMF (parameterised by concentration $\kappa$ rather than variance $\sigma^2$) is planned for the prototype implementation, alongside empirical comparison of vMF versus Gaussian-plus-renormalisation on routing accuracy.

## VIII. CONCLUSION

The Internet of Agents is materialising rapidly, supported by foundational standards (A2A, MCP), robust identity frameworks (DIDs, VCs), and enterprise-grade orchestration platforms. Yet a compliance deadlock persists: enterprises in financial services, healthcare, and defence cannot orchestrate AI agents across regulatory boundaries because existing protocols require the routing intermediary to access payload plaintext.

SS-ZKR resolves this deadlock through three interdependent mechanisms: differentially private blind routing that allows semantic capability matching without payload decryption; vector-weighted adaptive sanitisation with formal DP guarantees; and a spatial-to-cryptographic policy compiler that makes zero-knowledge security boundaries accessible to non-cryptographers. By design, SS-ZKR complements rather than competes with A2A, MCP, and DID/VC infrastructure. It adds the privacy-preserving routing layer that the emerging agent ecosystem currently lacks.

The immediate commercial impact is in regulated industries where multi-agent workflows are stalled by data sovereignty constraints. The longer-term impact is architectural: as the B2B agent economy emerges, privacy-preserving routing will be a prerequisite for competitive collaboration, letting organisations transact through shared agent infrastructure without exposing proprietary data to marketplace intermediaries.